\documentclass[a4paper, 11pt]{article}

\usepackage{graphicx}
\usepackage{amsmath}
\usepackage{amsfonts}
\usepackage{amssymb}
\usepackage{authblk}

\usepackage{graphicx}
\usepackage{braket}
\usepackage{physics}

\usepackage{tikz}
\usetikzlibrary{arrows,backgrounds,snakes,patterns}
\usetikzlibrary{shapes,arrows,chains}
\usepackage{verbatim}
\usepackage{booktabs}
\DeclareMathOperator{\sgn}{sgn}
\DeclareMathOperator{\diag}{diag}

\usepackage[flushleft]{threeparttable}

\usepackage{amsthm} 
\newtheorem{definition}{Definition}[section]
\newtheorem{theorem}{Theorem}
\newtheorem{corollary}{Corollary}

\usepackage{tikz}
\usetikzlibrary{arrows}
\usetikzlibrary{decorations.markings}
\usetikzlibrary{patterns}
\tikzset{
    thickest/.style={line width=3pt},
    empty/.style={decoration={markings,
    mark=at position #1 with {\fill[white,draw=black,thin] circle (3pt);}},postaction={decorate}},
    full/.style={decoration={markings,
    mark=at position #1 with {\fill circle (3pt);}},postaction={decorate}},
}

\date{\vspace{-5ex}}

\begin{document}

\title{Nuclear Numerical Range and Quantum Error Correction Codes for non-unitary noise models}

\author[1]{Patryk Lipka-Bartosik  \footnote{patryk.lipka.bartosik@gmail.com}}
\author[2,3]{Karol \.Zyczkowski}

\affil[1]{Faculty of Physics, University of Warsaw, ul. Pasteura 5, 02-093 Warsaw, Poland}
\affil[2]{Center for Theoretical Physics, Polish Academy of Sciences, \newline Al. Lotnik\'ow 32/44, 02-668 Warsaw, Poland}
\affil[3]{Institute of Physics, Jagiellonian University, ul. Reymonta 4, 30-059 Cracow, Poland}

\maketitle

\begin{abstract}
We introduce a notion of nuclear numerical range defined as the set of expectation values of a given operator $A$ among normalized pure states, which belong to the nucleus of an auxiliary operator $Z$. This notion proves to be applicable to investigate models of quantum noise with block-diagonal structure of the corresponding Kraus operators. The problem of constructing a suitable quantum error correction code for this model can be restated as a geometric problem of finding intersection points of certain sets in the complex plane. This technique, worked out in the case of two-qubit systems, can be generalized for larger dimensions.
\end{abstract}

\section{Introduction}
\label{intro}
Quantum information processing may potentially revolutionize classical computation based on ordinary bits. However, current constructions of universal quantum computer still cannot compete with classical computational machines. One of the main difficulties lies in the fact that systems on a quantum scale are extremely susceptible to any kind of external noise, as well as to erroneous action of quantum gates in a circuit. Therefore, to handle qubits effectively, there is a need for methods protecting quantum information against all possible disturbances. Two general approaches to this problem have been developed. The first one is based on the so-called \emph{decoherence-free subspaces} and exploits particular states of the Hilbert space that are immune to certain errors - a readable review of this methods can be found in Refs. \cite{lidar2003,lidar2014}. An alternative technique is based on \emph{quantum error-correcting codes} (QECC), which are quantum counterparts of the classical error-correcting codes. Quantum error-correcting codes are constructs which protect quantum information against some specified errors. This method of error-correction has been extensively studied in the case of unitary noise operations - see Ref. \cite{lidar2013} for a comprehensive introduction to this field. However, any real quantum operation can be only \emph{approximately} unitary and thus one has to consider non-unitary noise operations. Here the progess is slower than in the unitary case, mainly because of the increased complexity of the problem. For instance, in the unitary case, any product of two Kraus operators is normal, so its numerical range is determined by the spectrum, which is not longer true in the general case. The need for a constructive method of finding quantum error correction codes for non-unitary noise models provides a motivation toward this work. 

The main aim of this paper is to propose a method of finding quantum error correction codes for a class of non-unitary noise operators. The method, in the form that is presented here, can be effectively applied to noise models with short Kraus decomposition (consisting of \emph{two} operators). The main advantage of this method is that it allows to solve an algebraic problem (often untractable with other approaches) using an elementary geometrical construction. The paper is organized as follows. In Sect.~\ref{sec:1} we review some basic notions related to quantum error correction and specify the general form of a non-unitary quantum channel.  In Sect.~\ref{sec:2} we recall definitions of generalized numerical range, introduce the concept of \emph{nuclear numerical range} and present some of its properties. In Sect.~\ref{sec:3} we describe a geometric method of obtaining quantum error correction codes for block-diagonal Kraus operators. Precisely, using compression formalism based on the Knill-Laflamme conditions \cite{laflamme,bennet} and the notion of nuclear numerical range, we obtain projectors on the subspaces of the quantum error correction code. This is the main contribution of this work. In Sect.~\ref{sec:4} we provide two examples of block-diagonal channels and obtain the corresponding quantum error correction codes using method described in Sect.~\ref{sec:3}. We conclude this paper with a summary of results obtained and discuss possible generalizations of the method to higher-dimensional problems.
\section{Quantum Error Correction}
\label{sec:1}
In this section we recall the definition of Kraus operators and their role in description of noise in quantum systems. We invoke the \emph{Knill-Laflamme} conditions \cite{laflamme} for an error correction code of a particular noise model. In the end of this section we present a block-diagonal model of quantum noise that will be explored further in this work.

\subsection{Kraus representation of a quantum noise}
Consider a quantum system $\rho$ described in an $n$-dimensional Hilbert space $\mathcal{H}^{n}$. Assume that the system evolves according to some given error process (noise channel) represented by a superoperator $\Phi$ acting on  $\mathcal{H}^n$. According to the \emph{Kraus Representation Theorem} \cite{nielsen} any such superoperator can be written as the sum of $m$ matrix operators $A_i \in \mathcal{M}_n(\mathbb{C})$, where $\mathcal{M}_n(\mathbb{C})$ is the space of all complex-valued matrices of order $n$:
\begin{equation}
\label{eq:1}
\Phi[\cdot] = \sum_{i=1}^m A^{\dagger}_i[\cdot]A_i,
\end{equation}
where $\dagger$ denotes the adjoint operator. Matrices $A_i$ are called  \emph{Kraus operators}, and they satisfy the trace preserving condition: $\sum_{i=1}^{m} A_i^{\dagger} A_i = \mathbb{I}$. In this paper we consider (unless otherwise stated) models of noise acting on two-qubit systems described by two Kraus operators. The dimension of matrices $\{A_i\}$ is thus $n = 4$ and there are $m = 2$ of them. To solve the quantum error correction problem for a map $\Phi$, one has to look for subspaces $\mathcal{H}^n_C \subset \mathcal{H}^n$, which satisfy \emph{Knill-Laflamme Conditions} \cite{laflamme,bennet}. Error correcting code, labeled by $\Ket{\mathcal{C}}$, is itself a quantum state in subspace $\mathcal{H}^n_C$ of dimension $n$. We denote by $\{\Ket{\psi_i}\}$ a particular basis of this subspace (correction code basis), so that $\mathcal{H}^n_C = \text{span}\{\Ket{\psi_i}\}$ and by $P = \sum_i^n \Ket{\psi_i}\Bra{\psi_i}$ the projection operator on $\Ket{\mathcal{C}}$. According to \cite{laflamme} the following conditions are sufficient to reconstruct information about the system $\rho$ subjected to errors described by the set of $m$ Kraus operators $\{A_i\}$:
\begin{align}
\label{eq:2}
PA^{\dagger}_iA_jP = \lambda_{ij} P \quad \text{for} \quad i,j = 1, 2, \ldots m.
\end{align} 
where $\lambda_{ij}$ are called \emph{compression values} of the error correcting code. The problem of determining the projectors $P$ is related to an algebraic \emph{compression problem} \cite{choi2006}. In our case the invariant subspace of code, $\mathcal{H}^n_C$, is two-dimensional, so $P$ is a projector on a two-dimensional subspace ($P = P_2$). Denoting $T_{ij} \equiv A^{\dagger}_iA_j$ we can write:
\begin{align}
\label{eq:3}
P_2 T_{ij} P_2 = \lambda_{ij} P_2, \quad i,j = 1, 2, \ldots, m
\end{align}
Determining quantum error-correction code for the error model described by the set of Kraus operators $\{A_i\}$ is equivalent to finding a subspace $P_2$ that satisfies the above set of equations.
\subsection{Noise models with block-diagonal Kraus operators}
The problem we adress in this paper involves finding correction subspace $\mathcal{H}^n_C$, which amounts to determining projections $P_2$ from the last equation. The main difficulty lies in the fact that $P_2$ has to satisfy the Knill-Laflamme conditions (\ref{eq:3})  for all operators $T_{ij}$ simultaneously. In what follows we consider the Kraus operators with a block-diagonal structure. It will be convenient to introduce a short-hand notation:
\begin{align}
\label{eq:4}
T_{11} &= A_1^{\dagger} A_1 = E_{11} \oplus F_{11} = \begin{bmatrix}
E_{11} & 0  \nonumber \\
0 & F_{11}
\end{bmatrix}, \\ 
T_{12} &= A_1^{\dagger} A_2 = E_{12} \oplus F_{12} =
\begin{bmatrix} 
E_{12} & 0 \\
0 & F_{12}, 
\end{bmatrix}, \\ \nonumber
T_{22} &= A_2^{\dagger} A_2 = \mathbb{I} - T_{11}, \qquad T_{21} = A_2^{\dagger} A_1 = T_{12}^{\dagger},
\end{align}
where  $E_{11}, E_{12}, F_{11}, F_{12} \in \mathcal{M}_2(\mathbb{C})$. Note that if Eq. (\ref{eq:3}) is fulfilled for $T_{12}$, it is also fulfilled for the adjoint $T_{12}^{\dagger} = T_{21}$. Moreover, due to the normalization condition we can write $T_{22} = \mathbb{I}-T_{11}$. Thus, we effectively look for simultaneous solutions of the compression problem for two operators $T_{11}$ and $T_{12}$. There are several approaches of solving problems of this kind if matrices $T_{ij}$ are normal (see for example the \emph{eigenvector-pairing method} introduced in \cite{choi}). Here we consider a broader range of models of non-unitary Kraus operators, for which matrices $T_{ij}$ need not be normal. Thus, the techinques developed in literature cannot be applied here in a straightforward manner. In Sect.~\ref{sec:3} we will develop a new method for solving this type of problems. 

\section{Mathematical Tools}
\label{sec:2}
In this section we review the notion of \emph{higher order numerical range} and introduce the concept of nuclear numerical range. We state several its properties that will be further explored to determine the correction code subspaces for models considered in Sect.~\ref{sec:3} and \ref{sec:4}.

\subsection{Higher order numerical range}
The algebraic form of equation (\ref{eq:3}) suggests that the problem can be approached using the so-called \emph{rank-}$k$ numerical range  $\Lambda_k(T)$ of matrix $T$, introduced in \cite{choi},
\begin{align}
\label{eq:5}
\Lambda_k(T) = \{ \lambda \in \mathbb{C}: P T P = \lambda P \text{ for some } P \in \mathcal{P}_k \},
\end{align}
where $\mathcal{P}_k$ is the set of all rank-$k$ projections on space $\mathcal{H}^n$. Unit vectors which yield compression value $\lambda = \lambda_0$ are called \emph{generating vectors} (or generators) of $\lambda_0$.  The two special cases ($k = 1$ and $k = 2$) are of particular interest in this paper. Notice that setting $k = 1$ yields the standard numerical range $\Lambda_1(T)$, often denoted by $W(T)$ \cite{horn}:
\begin{align}
\label{eq:6}
\Lambda_1(T) = W(T) = \{ \lambda \in \mathbb{C}: \lambda = \Braket{\psi|T|\psi},\, \Braket{\psi|\psi} = 1 \}.
\end{align}
On the other hand, the case $k = 2$ gives the numerical range of rank two:
\begin{align}
\label{eq:7}
\Lambda_2(T) = \{\lambda \in \mathbb{C}: P_2 T P_2 = \lambda P_2,\, P_2 = P_2^{\dagger} \text{ and } P_2P_2=P_2  \}, \nonumber
\end{align}
Based on the definition of higher rank numerical range $\Lambda_k$, one can derive the following properties \cite{choi} :
\begin{description}
\item[(P1)] For any $a,\, b \in \mathbb{C}$, $\Lambda_k(aA+b I) = a \Lambda_k(A) + b.$
\item[(P2)] For any unitary $U \in \mathcal{M}_n$, $\Lambda_k(U^{\dagger} A U) = \Lambda_k(A).$
\item[(P3)] If A is Hermitian and $\{ \lambda_i\}$ is an ordered set of eigenvalues of a matrix $A$, then $\Lambda_k(A) = [\lambda_{n-k+1}(A), \lambda_k(A)]$, where interval is an empty set if $\lambda_{n-k+1}(A) > \lambda_k(A)$ when $k > n/2$. 
\item[(P4)] For any $k_1,\, k_2 \in \mathbb{N}$, $\Lambda_{k_1}(A) \cap \Lambda_{k_2}(B) \subseteq \Lambda_{k_1+k_2}(A\oplus B)$.  
\end{description}
Having defined higher order numerical range, we are ready to introduce a new structure, the nuclear numerical range.

\subsection{Nuclear numerical range}
In this section we define a structure that will prove useful to determine error-correction subspaces.
\begin{definition}
\label{d1}
Given two square, complex-valued matrices of the same size: $A$, $Z \in \mathcal{M}_n(\mathbb{C})$, the $Z$-nuclear numerical range of $A$, labeled by $W(A|Z)$, is defined as the set of expectation values of $A$ among pure states which belong to the kernel of $Z$, 
\begin{align}
W(A|Z) = \{ z \in \mathbb{C}: z = \Braket{\psi|A|\psi},\, \Braket{\psi|\psi} = 1  \text{ and }\Braket{\psi|Z|\psi} = 0\}.
\end{align}
\end{definition}
By definition, the nuclear numerical range is a set in the complex plane contained in $W(A)$, that is $W(A|Z) \subseteq W(A)$ for all matrices $Z$. Note that the notion of nuclear numerical range $W(A|Z)$ belongs to the class of \emph{restricted numerical ranges} described in \cite{gawron}, since the set of states $\Ket{\psi}$ used in Definition \ref{d1} is restricted to the set containing the kernel (\emph{nucleus}) of $Z$. The standard numerical range $W(A)$ is convex, while the set formed by $Z$-nuclear numerical range is not convex in general. In this paper we mostly consider $W(A|Z)$ for matrices of size $n = 2$. Let us now present some properties of nuclear numerical range. 

Let $X,\,Y$ be two non-zero matrices in $\mathcal{M}_n(\mathbb{C})$ and $U$ be some unitary matrix in $\mathcal{M}_n(\mathbb{C})$. Then, the following properties follow directly from the definition of $W(A|Z)$:
\begin{description}
\item[(N1)] $W(A|0) = W(A)$ is the standard numerical range. 
\item[(N2)] $W(A|X^{\dagger} X) = \varnothing$ for any nonsingular $X$ and all matrices $A$. 
\item[(N3)] $W(A|A-\alpha \mathbb{I}) = \{ \alpha \}$, $\alpha \in W(A)$.
\item[(N4)] $W(A|Z) = W(A|\,(-1) \cdot Z)$.
\item[(N5)] $W(X|X-Y) = W(Y|Y-X)$.
\item[(N6)] $W(A + \alpha Z|Z) = W(A|Z)$. 
\item[(N7)] If $Z$ is traceless, then $W(A|Z)$ is not empty.
\end{description}
Moreover, the following properties hold for two dimensional matrices ($n = 2$). Let us assume that $Z$ is a hermitian matrix and let $\nu_1$ and $\nu_2 > \nu_1$ denote its eigenvalues. Then:
\begin{description}
\item[(N8)] $W(A|Z)$ forms a (possibly degenerate) elliptic curve.
\item[(N9)] If $\sgn(\nu_1) \neq \sgn(\nu_2)$, then $W(A|Z)$ is not empty.
\item[(N10)] If $Z$ is normal and $\lambda \in [\nu_1, \nu_2]$, then $W(A|Z-\lambda \mathbb{I}) = W(A)$. 
\end{description}
The proof of (N$1)-($N$6$) is straightforward from the definition of $W(A|Z)$ and properties (P$1)-($P$2$) from previous subsection. To prove property (N$7$) note that $W(A|Z)$ always contains the point $\frac{1}{2}\Tr A$. Thus, $\frac{1}{2}\Tr A = 0 \in W(Z)$ and $W(A|Z)$ is not empty. In order to prove property (N$8$) we use the following theorem, proved in Appendix A.
\begin{theorem}
\label{t1}
Let $Z$ be a normal matrix of order two with real-valued entries. Furthermore, let $A$ be an arbitrary complex-valued matrix of order two. Consider:
\begin{align*}
Z &= 
\begin{bmatrix}
2 a & b \\
b & 2 c
\end{bmatrix} \in \mathcal{M}_2^{\mathbb{R}}, \quad
A =
\begin{bmatrix}
d & f \\
g & h
\end{bmatrix} \in \mathcal{M}_2^{\mathbb{C}}.
\end{align*}
Then, there exists a set of normalized states $\{\Ket{\psi}\}$, $\Ket{\psi} \in \mathbb{C}^2$, parametrized by a phase $\varphi \in [0, 2\pi)$ and a real number $\lambda \in W(Z)$, which satisfies the following set of simultaneous equations:
\begin{align}
\label{teq:1}
\begin{cases}
\Braket{\psi|Z|\psi} = \lambda,\\
\Braket{\psi|A|\psi} = z(\varphi, \lambda),
\end{cases}
\end{align}
where $z(\varphi, \lambda)$ forms an elliptic disk in the complex plane parametrized by $\varphi$ and $\lambda$:
\begin{align}
\label{teq:2}
z(\varphi, \lambda) = z_0 + w \lambda + p(\lambda) [q \cos \varphi + r \sin \varphi].
\end{align}
Variables $z_0$, $w$, $p(\lambda)$, $q$ and $r$ are defined in Eq. (\ref{leq:4}). The family of states $\{\Ket{\psi}\}$ is given by:
\begin{align}
\Ket{\psi} = 
\begingroup
\renewcommand*{\arraystretch}{1.3}
\begin{pmatrix}
\cos \alpha \cos \frac{\theta}{2} + \text{e}^{i \varphi} \sin \alpha \sin \frac{\theta}{2} \\ 
\sin \alpha \cos \frac{\theta}{2} + \text{e}^{i \varphi} \cos \alpha \sin \frac{\theta}{2},
\end{pmatrix}
\endgroup
\end{align}
where $\tan 2\alpha = b / (a-c)$ and $\cos \theta = (\lambda - a - c) / \sqrt{b^2 + (a-c)^2}$.
\end{theorem}
From above theorem we can deduce the following simple corollary:
\begin{corollary}
\label{c1}
$W(A|Z - \lambda \mathbb{I}) = z(\varphi, \lambda)$, where $\varphi \in [0, 2\pi)$ and $z(\varphi, \lambda)$ is given by Eq. (\ref{teq:2}).
\end{corollary}
The proof of property (N$8$) follows directly from Theorem \ref{t1} if we take $\lambda = 0$. In order to prove property (N$9$)  note that by the elliptic range theorem \cite{li}, the set $W(Z)$ forms an elliptic disk with foci at eigenvalues of $Z$, that is $\nu_1$ and $\nu_2$. By the convexity property, $W(Z)$ must also contain a line with endpoints $(\nu_1, \nu_2)$. Since $\sgn(\nu_1) = -\sgn(\nu_2)$, the line either passes through the origin or is a singular point at the origin. Thus, there exists $\Ket{\psi}$ such that $\Braket{\psi|Z|\psi} = 0$ and $W(A|Z)$ is not an empty set. Property (N$10$) can be proven after noticing that if one allows $\lambda$ to take all possible real values between the two eigenvalues of $Z$, then there is no restriction on vector $\Ket{\psi}$ (cosine of azimuthal angle given by Eq. (\ref{leq:2}) in the Appendix takes all real values between $-1$ and $1$). This means that $W(A|Z-\lambda\mathbb{I})$ is the full numerical range $W(A)$.
\\ \\
It is worth to emphasise that the unitary invariance of standard numerical range does not hold in the case of nuclear numerical range, that is $W(A|UZU^{\dagger}) \neq W(A|Z)$. It can be easly seen by considering:
\begin{align}
A = \begin{bmatrix}
a & b \\
c & d
\end{bmatrix}, \quad
Z = \begin{bmatrix}
0 & 2 \\
0 & 0
\end{bmatrix}, \quad 
U = \frac{1}{\sqrt{2}}\begin{bmatrix}
1 & 1 \\
1 & -1
\end{bmatrix}.
\end{align} 
Computing $W(A|Z)$ yields a single point $\lambda = a \in \mathbb{C}$, whereas $W(A| UZU^{\dagger}) = \lambda' = (a+b+c+d)/2 \in \mathbb{C}$. Thus, for a general matrix $A$, $\lambda \neq \lambda'$. 
Similary we can find $W(UAU^{\dagger}|Z) \neq W(A|Z)$.

\section{Determination of code subspaces for Kraus operators with block-diagonal structure}
\label{sec:3}
In this section we describe a method of constructing projectors $P_2$ onto the code subspace $\mathcal{H}^n_C$ for not-normal matrices. Let us now return to the basic problem and recall equation (\ref{eq:3}). Because of the block-diagonal structure of $T_{ij}$ we expect that $P_2$ will have a similar block-diagonal structure. We emphasize that this choice of $P_2$ is not the most general possible and reduces the total set of possible correction codes we can obtain. Let us call the upper and lower of the blocks of $P_2$ by $P_E$ and $P_F$, respectively. This allows us to write:
\begin{align}
\label{eq:9}
&P_{2} = P_E \oplus P_F = \begin{bmatrix}
P_{E} & 0 \\
0 & P_{F}
\end{bmatrix}.
\end{align}
Using property (P$4$) and setting $k_1 = k_2 = 2$ we can reduce the initial problem of determining one $4$-by-$4$ projection matrix $P_2$ to a problem of finding two $2$-by-$2$ projection matrices $P_E$ and $P_F$. Writing explicitly: 
\begin{align}
\label{eq:10}
\Lambda_{2}(T_{ij}) = \Lambda_{2} (E_{ij} \oplus F_{ij}) \supseteq \Lambda_{1}(E_{ij}) \cap \Lambda_{1}(F_{ij}).
\end{align}
We can rewrite Eq. (\ref{eq:3}) in the matrix notation:
\begin{align}
\label{eq:11}
P_2T_{ij}P_2 = 
\begin{bmatrix}
P_{E}E_{ij}P_E & 0 \\
0 & P_FF_{ij}P_F
\end{bmatrix}  = \begin{bmatrix}
\lambda_{ij}^E P_E & 0 \\
0 & \lambda_{ij}^F P_F
\end{bmatrix}.
\end{align}
By definition, $\lambda_{ij}^E \in \Lambda_1(E_{ij})$ and $\lambda_{ij}^F \in \Lambda_1(F_{ij})$ are points in respective standard numerical ranges. The above equality is satisfied only if the condition $\lambda_{ij}^E = \lambda_{ij}^F = \lambda_{ij}$ holds for all values of $i$ and $j$. This means that both points, $\lambda_{ij}^E$ and $\lambda_{ij}^F$ lie in the intersection $\Lambda_{1}(E_{ij}) \cap \Lambda_{1}(F_{ij})$ for every choice of $i,j$. In order to determine $P_E$ and $P_F$ (which are projections onto points in that intersection) note that we can rewrite them in respective correction code bases as:
\begin{align}
\label{eq:12}
P_E = \Ket{\psi_E}\Bra{\psi_E} \qquad \text{and} \qquad P_F = \Ket{\psi_F}\Bra{\psi_F},
\end{align} 
where both vectors are normalized $\Braket{\psi_E|\psi_E} = \Braket{\psi_F|\psi_F} = 1$. Note that the states $\Ket{\psi_E}$ and $\Ket{\psi_F}$ are both generators of the set of compression values $\lambda_{ij}$, for $i, j = 1, 2$. The intersections that we are interested in, $\Lambda_{1}(E_{ij}) \cap \Lambda_{1}(F_{ij})$, are sets of points for which the following statement holds: 
\begin{align}
\label{eq:13}
\Braket{\psi_E|E_{ij}|\psi_E} = \lambda_{ij} = \Braket{\psi_F|F_{ij}|\psi_F}.
\end{align}
Due to the structure assumed in Eq. (\ref{eq:4}) not all of the above equations are independent. Having this in mind, we are left with the following set of two equations:
\begin{align}
\label{eq:14}
\begin{cases}
\Braket{\psi_E|E_{11}|\psi_E} &= \lambda_{11} = \Braket{\psi_F|F_{11}|\psi_F}, \ \\
\Braket{\psi_E|E_{12}|\psi_E} &= \lambda_{12} = \Braket{\psi_F|F_{12}|\psi_F}.
\end{cases}
\end{align}
To solve this set of equations we use the concept of nuclear numerical range discussed in Sect.~\ref{sec:3}. Consider the following two sets:
\begin{align*}
W(E_{12}|E_{11} - \lambda_{11} \mathbb{I}) \qquad \text{and} \qquad  W(F_{12}|F_{11} - \lambda_{11} \mathbb{I}).
\end{align*}
Notice that points in the complex plane which satisfy Eqs. (\ref{eq:14}) are exactly the ones that constitute the intersection of above nuclear numerical ranges. Let us label this intersection by $\Gamma (\lambda_{11})$:
\begin{align}
\label{eq:15}
\Gamma(\lambda_{11}) =  W(E_{12}|E_{11} - \lambda_{11} \mathbb{I}) \cap W(F_{12}|F_{11} - \lambda_{11} \mathbb{I}) \subset \mathbb{C}.
\end{align}
Clearly, $\Gamma(\lambda_{11}) \subset W(E_{12}) \cap W(F_{12})$. Since both matrices $E_{11}$ and $F_{11}$ are normal, by property (N$8$) we conclude that for a given value of $\lambda_{11}$ these two sets are elliptic curves in the complex plane. If we treat $\lambda_{11}$ as a parameter whose range is the appropriate line segment $W(E_{11})$, then by property (N$10$) we have $W(E_{12}|E_{11} - \lambda_{11} \mathbb{I}) = W(E_{12})$. Similar statement holds for a pair $F_{12}$ and $F_{11}$. In our problem, in order to satisfy Eq. (\ref{eq:14}), the number $\lambda_{11}$ must be contained in both $W(E_{11})$ and $W(F_{11})$. Thus $\lambda_{11}$ must be contained in the intersection $W(E_{11}) \cap W(F_{11})$, which is simply a line segment. If we denote the sets of eigenvalues of matrices $E_{11}$ and $F_{11}$ by $\{\nu_{E}\}$ and $\{\nu_{F}\}$ respectively, then in order to fulfill the first condition from Eq. (\ref{eq:14}), one has to satisfy:
\begin{align}
\label{eq:15a}
\max\limits_{ \{\nu_{E}\} } > \lambda_{11} > \min \limits_{ \{\nu_{F}\} } \quad \text{or} \quad 
\min\limits_{ \{\nu_{E}\} } < \lambda_{11} < \max \limits_{ \{\nu_{F}\} }.
\end{align}   
Let us label by $\Omega$ the set of all allowable values of $\lambda_{11}$. Then, the set $\Omega $ is not empty if and only if:
\begin{align}
\label{eq:15b}
\Tr E_{11} \pm \left[\left(\Tr E_{11}\right)^2-4\det E_{11} \right]^{1/2} \gtrless
\Tr F_{11} \mp \left[\left(\Tr F_{11}\right)^2-4\det F_{11} \right]^{1/2},
\end{align}  
where the symbol $\gtrless$ corresponds to $+$ and $-$ signs respectively, so Eq. (\ref{eq:15b}) contains two inequalities. The above equation follows from the fact that the two eigenvalues of a $2$-by-$2$ matrix A are given by a formula $\frac{1}{2} \Tr A  \pm \frac{1}{2}\sqrt{(\Tr A)^2 - 4\det A}$. By Corollary \ref{c1} we can conclude that $W(E_{12}|E_{11} - \lambda_{11}) = z(\varphi_E, \lambda_{11})$ and $W(F_{12}|F_{11} - \lambda_{11}) = z(\varphi_F, \lambda_{11})$, where function $z(\varphi, \lambda)$ is defined in Eq. (\ref{teq:2}). The intersection $\Gamma(\lambda_{11})$ is determined by:
\begin{align}
\label{eq:15c}
\Gamma(\lambda) = \{z \in \mathbb{C}|z = z(\varphi_E, \lambda_{11}) = z(\varphi_F, \lambda_{11}),\  \varphi_E, \varphi_F\in [0, 2 \pi), \ \lambda_{11} \in \Omega\}.
\end{align}
In order to determine the vectors $\Ket{\psi_E}$ and $\Ket{\psi_F}$, which are useful to construct projectors on the respective correction code bases (recall Eq. (\ref{eq:12})), one can in principle find points in $\Gamma(\lambda_{11})$ determined by $\lambda_{11} \in \Omega$ and their phase angles $\varphi_E$ and $\varphi_F$. Following the proof of Theorem \ref{t1}, one diagonalizes $E_{11}$ and $F_{11}$ using orthogonal matrices $U_E$ and $U_F$:
\begin{align}
\label{eq:16}
U_E = \begin{bmatrix}
\cos \alpha_E & \sin \alpha_E \\
\sin \alpha_E & - \cos \alpha_E
\end{bmatrix}, \qquad
U_F = \begin{bmatrix}
\cos \alpha_F & \sin \alpha_F \\
\sin \alpha_F & - \cos \alpha_F
\end{bmatrix}.
\end{align}
Now one introduces the transformed states: $\Ket{\psi_E} = U_E \Ket{\phi_E}$ and $\Ket{\psi_F} = U_F \Ket{\phi_F}$, where $\Ket{\phi_E}$ and $\Ket{\phi_F}$ are given by: 
\begin{align}
\label{eq:17}
\Ket{\phi_E}=
\begin{pmatrix}
\cos\frac{\theta_E}{2} \\ \text{e}^{i \varphi_E}\sin\frac{\theta_E}{2}
\end{pmatrix}, \qquad 
\Ket{\phi_F}=
\begin{pmatrix}
\cos\frac{\theta_F}{2} \\ \text{e}^{i \varphi_F}\sin\frac{\theta_F}{2}
\end{pmatrix},
\end{align}
where $i = \sqrt{-1}$ denotes the complex imaginary unit. In order to determine these projection vectors one has to find angles $\theta_{\sigma}, \varphi_{\sigma}, \alpha_{\sigma}$ for $\sigma \in \{E, F\}$ in terms of a point $\widetilde{z} = \widetilde{x} + i \widetilde{y} \in \Gamma(\lambda)$ in the complex plane and a real parameter $\lambda$. Let us suppose that we have found a point $\widetilde{z} \in \Gamma(\lambda)$ for some value of $\lambda$. We shall consider only the case for matrices $E_{11}$ and $E_{12}$ since for the matrices $F_{11}$ and $F_{12}$ the same pattern can be applied. Our current task is to find the projection vectors $\Ket{\varphi_E}$. Using Eq. (\ref{leq:2}) one can determine the azimuthal angle for projection $\theta_E$ in terms of $\lambda$: 
\begin{align}
\label{eq:18}
\cos \theta_{E} = \frac{1}{\epsilon_E}\left(\lambda - \frac{1}{2} \Tr E_{11}\right),
\end{align}
where $\pm \epsilon_E$ denotes the eigenvalues of matrix $E_{11}$ (defined in Appendix A, in the text above Eq. \ref{leq:1}). Using the definition of $\widetilde{z} = \widetilde{z}(\varphi, \lambda) \in \Gamma(\lambda)$ given in Eq. (\ref{leq:5}) with the following substitution: $(A, Z) \rightarrow (E_{12}, E_{11})$, one finds:
\begin{align*}
x(\varphi_E) = \frac{\widetilde{x} - x_0(\lambda)}{p(\lambda)}, \qquad y(\varphi_E) = \frac{\widetilde{y} - y_0(\lambda)}{p(\lambda)}.
\end{align*}
Using the above expressions and Eqs. (\ref{leq:6a}) and (\ref{leq:6b}) one can determine the polar angle $\varphi_E$,
\begin{align}
\label{eq:19}
\cos \varphi_E = \frac{x r_2 - y r_1}{q_1 r_2 - r_1 q_2} = \frac{1}{p(\lambda)}\frac{r_2\left[\widetilde{x} - x_0(\lambda)\right] - r_1\left[\widetilde{y} - y_0(\lambda))\right]}{q_1 r_2 - r_1 q_2}.
\end{align} 
The angle $\alpha_E$ can be computed by recalling that $\tan 2\alpha_E$ is related by Eq. (\ref{leq:1aa}) to elements of matrix $E_{11}$. The family of states $\Ket{\psi_E}$ is then given by:
\begin{align}
\label{eq:20}
\Ket{\psi_E} = 
\begingroup
\renewcommand*{\arraystretch}{1.3}
\begin{pmatrix}
\cos \alpha_E \cos \frac{\theta_E}{2} + \text{e}^{i \varphi_E} \sin \alpha_E \sin \frac{\theta_E}{2} \\ 
\sin \alpha_E \cos \frac{\theta_E}{2} + \text{e}^{i \varphi_E} \cos \alpha_E \sin \frac{\theta_E}{2}
\end{pmatrix}.
\endgroup
\end{align}  
In a similar manner we can determine the projection vectors $\Ket{\psi_F}$ for matrices $F_{11}$ and $F_{12}$. Thus, our initial problem of finding error correction code and solving the compression Eqs. (\ref{eq:3}), equivalent to the set of Eqs. (\ref{eq:14}) for projectors $\Ket{\psi_E}$ and  $\Ket{\psi_F}$, is reduced into a geometric problem of finding intersection points of two elliptic curves in the complex plane. The quantum code subspace - the projection $P_2$ from Eq. (\ref{eq:3}) - is then given by a matrix of order four, given by the direct sum of two projectors of size two:
\begin{align}
\label{eq:21}
P_2 = \Ket{\psi_E} \Bra{\psi_E} \oplus \Ket{\psi_F} \Bra{\psi_F}.
\end{align}

\section{Exemplary Quantum Error Correction Codes}
\label{sec:4}
In this section we present two examples of non-unitary quantum channels and determine their quantum error correction code subspaces $\mathcal{H}^n_C$ using the method described in the previous section.

\subsection{Simplified two-qubit amplitude-damping channel}
The amplitude-damping channel (AD channel) is an important channel describing effects due to loss of energy of a quantum system \cite{nielsen}. Here we consider two-level systems (qubits), but channels describing arbitrary $n$-level systems are also known \cite{grassl}. Moreover, an interesting study of generalized amplitude-damping channels based on approximate quantum error-correction schemes appeared recently \cite{cafaro}. Exemplary physical processes which can be described by this channel include the relaxation of atom from its excited state to the ground state \cite{audretsch}, sending a quantum state from one location to another using a spin chain \cite{giovanetti,bose} and attenuation of a photon in a cavity \cite{hongyi}. The Kraus representation of one-qubit amplitude damping channel acting on state $\rho$ with probability $p$, where $0 \leq p \leq 1$, is given by:
\begin{align*}
\Phi^{1AD}(\rho) = B_1(p) \rho B_1^{\dagger}(p) + B_2(p) \rho B_2^{\dagger}(p),
\end{align*}
where the Kraus operators $B_1(p)$ and $B_2(p)$ are defined:
\begin{align*}
B_1(p) = \begin{bmatrix}
1 & 0 \\ 0 & \sqrt{1-p}
\end{bmatrix},\qquad
B_2(p) = \begin{bmatrix}
0 & \sqrt{p} \\ 0 & 0
\end{bmatrix}.
\end{align*}
To extend this channel into the two-qubit system one can consider a product of two one-qubit channels:
\begin{align*}
\Phi^{2AD} = \Phi_1^{1AD} \otimes \Phi_2^{1AD}.
\end{align*}
This channel is given by four Kraus operators. The bi-partite channel will be described by a sum of four new Kraus operators (all possible tensor products of $B_1$ and $B_2$) acting on a two-qubit state $\rho_1 \otimes \rho_2$. Assuming that the damping occurs with probability $p_1$ on the first qubit and $p_2$ on the second, we may write:
\begin{align*}
C_{ij} = B_i(p_1)\otimes B_j(p_2), \qquad 1 \leq i, j \leq 2.
\end{align*}
To simplify the model we consider a two-qubit channel defined by two Kraus operators: $A_1 = \sqrt{\mathbb{I} - A_2^{\dagger} A_2}$ and $A_2 = C_{12} = B_1 (p_1) \otimes B_2 (p_2)$, where $A_1$ is defined up to a unitary transformation. We make the following choice of Kraus operators:
\begin{align*}
A_1 = 
\setlength{\arraycolsep}{4pt}
\begin{bmatrix}
0 & \sqrt{1-p_2} & 0 & 0 \\
1 & 0 & 0 & 0 \\
0 & 0 & 1 & 0 \\
0 & 0 & 0 & \sqrt{1-p_2(1-p_1)} \end{bmatrix}
,\qquad A_2 = \begin{bmatrix}
0 & \sqrt{p_2} & 0 & 0 \\
0 & 0 & 0 & 0 \\
0 & 0 & 0 & \sqrt{p_2(1-p_1)} \\
0 & 0 & 0 & 0 \end{bmatrix}.
\end{align*}
The trace-preserving channel analyzed here is then given by: 
\begin{align}
\label{eq:ex11}
\Phi^{2AD}(\rho_1 \otimes \rho_2) = A_{1}^{\dagger}(\rho_1 \otimes \rho_2)A_{1} + A_{2}^{\dagger}(\rho_1 \otimes \rho_2)A_{2}. 
\end{align}
One can determine the quantum error-correction code for this channel by solving the compression problem given in Eq. (\ref{eq:3}). Let us solve it using the geometrical method presented in Sect.~\ref{sec:3}. In the first step we compute matrices $T_{11}$ and $T_{12}$ given in Eq. (\ref{eq:4}), from which we obtain matrices $E_{ij}$ and $F_{ij}$ for $1 \leq i, j \leq 2$:
\begin{equation*}
\setlength{\arraycolsep}{5pt}
\begin{split}
E_{11} &= \begin{bmatrix} 
1&0\\0&1-p_2 
\end{bmatrix}, \\
E_{12} &= \begin{bmatrix} 
0&0\\0&\sqrt{p_2(1-p_2)}
\end{bmatrix},
\end{split}
\qquad
\begin{split}
F_{11} &= \begin{bmatrix} 
1&0\\0&1-p_2(1-p_1)
\end{bmatrix}, \\
F_{12} &= \begin{bmatrix} 
0& \sqrt{p_2(1-p_1)}\\0&0
\end{bmatrix}.
\end{split}
\end{equation*}
Our aim is to find projection operator $P_2 = P_E \oplus P_F$ from Eq. (\ref{eq:11}), where $P_E$ and $P_F$ are given by (\ref{eq:12}), which is equivalent to set of equations (\ref{eq:14}). To find these operators we use notion of nuclear numerical range. We first compute the intersection $\Gamma(\lambda_{11})$, where $\lambda_{11} \in \Omega = W(E_{11})\cap W(F_{11})$ and from this intersection we determine projection operators $P_E$ and $P_F$. 
\begin{figure}[!ht]
  \label{fig:4}
  \centering
  \begin{tikzpicture}
   
	\begin{scope}[xshift=10cm]
		\draw[thick,->,black!50] (-1.0,0) -- (10.0,0);
        
		\draw[thick] (1.5, -0.12) -- (1.5, 0.12);
      	\node at (1.5,-0.4) {$1-p_2$};	
      	
      	\draw[thick] (4.5, -0.12) -- (4.5, 0.12);
      	\node at (4.5,-0.4) {$1-p_2(1-p_1)$};	
      	
      	\draw[thick] (7.5, -0.12) -- (7.5, 0.12);
      	\node at (7.5,-0.4) {};	
      	
      	\draw[thick] (9, -0.12) -- (9, 0.12);
      	\node at (9,-0.4) {$1$};

       	\draw[thickest,full=0,full=1] (1.5,.5) -- (9,.5);
       	\draw[thickest,full=0,full=1] (4.5,-0) -- (9,-0);
      	
      	\draw[dashed] (1.5,0) -- (1.5,1);
       	\draw[dashed] (4.5,0) -- (4.5,1);
       	\draw[dashed] (9,0) -- (9,1);
       	
       	\draw[pattern=north west lines, pattern color=blue, solid, very thick] (4.5,0) rectangle (9,.5);

		\node[above right] (A) at (4.5, .5) {$W(E_{11})$};
		\node[above right] (B) at (6.1, -0.75) {$W(F_{11})$};
		
		\node[above right] (C) at (1.5, 1.5) {$\Omega=W(E_{11})\cap W(F_{11})$};
		\node[above right] (D) at (6.65, 0) {};
		\draw[->, to path={-| (\tikztotarget)}, thick]
  			(C) edge (D) ;

	\end{scope}
	
	\node[xshift=10cm] (X) at (9.8,0.3) {$Re$};

\end{tikzpicture}
  \caption{Standard numerical ranges $W(E_{11})$ and $W(F_{11})$ for the noise model given by (\ref{eq:ex11}) form line segments on the real axis. For clarity the range $W(E_{11})$ is lifted up. Their intersection forms the set $\Omega$, which determines all possible values of the parameter $\lambda_{11}$.}
\end{figure}
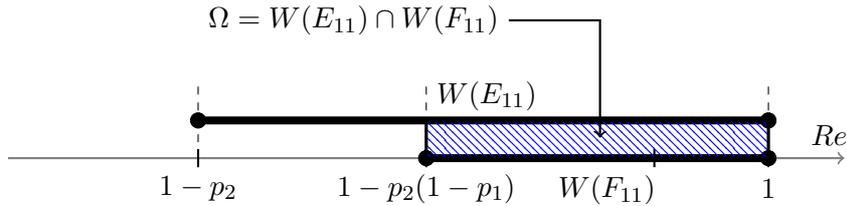

Let us introduce kets $\Ket{\psi_E} = U_E \Ket{\phi_E}$ and $\Ket{\psi_F} = U_F \Ket{\phi_F}$, where diagonalizing unitary matrices $U_E$ and $U_F$ are given by Eq. (\ref{eq:16}). Since $E_{11}$ and $F_{11}$ are already diagonal we may assume that $U_E = U_F = \mathbb{I}$ and thus $\Ket{\psi_E} = \Ket{\phi_E}$ and $\Ket{\psi_F} = \Ket{\phi_F}$, where $\Ket{\phi_E}$ and $\Ket{\phi_F}$ are yet arbitrary and parametrized according to Eq. (\ref{eq:17}). In order to find the azimuthal angles $\theta_{E}$ and $\theta_{F}$ one has to assure that the set $\Omega$ is not empty, which means that the overlapping condition (\ref{eq:15a}) holds. In this case this condition reduces to the following two expressions:
\begin{align*}
p_1 \leq 1 \quad \text{and} \quad p_2 \geq 0.
\end{align*}
Both of above conditions are satisfied by $p_1$ and $p_2$. We can now write respective standard numerical ranges as:
\begin{align*}
W(E_{11}) = 1 - \frac{p_2}{2} \left(1 - \cos \theta_E\right), \qquad W(F_{11}) =1 - \frac{p_2}{2}(1-p_1)(1-\cos\theta_F).
\end{align*}
This allows us to conclude that the set $\Omega$ is given by the intersection of two sets, $\Omega = W(E_{11})\cap W(F_{11}) = \bigl[ 1-(1-p_1)p_2, 1\bigr]$, as shown in Fig. 1. If we now treat $\theta_E$ and $\theta_F$ as functions of the parameter $\lambda_{11}\in \Omega$, we get:
\begin{align}
\label{eq:ex01}
\cos \theta_E = 1 - \frac{2}{p_2}(1-\lambda_{11}) \quad \text{and} \quad \cos \theta_F = 1 - \frac{2}{p_2 (1-p_1)} \left(1 -\lambda_{11}\right).
\end{align}
In order to find $W(E_{12}|E_{11}-\lambda_{11}\mathbb{I})$ and $W(F_{12}|F_{11}-\lambda_{11}\mathbb{I})$ we first compute the standard numerical range of matrices $E_{12}$ and $F_{12}$:
\begin{align*}
W(E_{12}) &= \Braket{\phi_E|E_{12}|\phi_E} = \frac{1}{2} \sqrt{p_2(1-p_2)} (1 - \cos \theta_E),\\
W(F_{12}) &= \Braket{\phi_F|F_{12}|\phi_F} = \frac{1}{2} e^{i \varphi_F} \sqrt{p_2(1-p_1)} \sin \theta_F.
\end{align*}
Using Eqs. (\ref{eq:ex01}) in the above expressions we obtain respective nuclear numerical ranges:
\begin{align*}
W(E_{12}|E_{11}-\lambda_{11} \mathbb{I}) &= (1-\lambda_{11})\sqrt{\frac{1}{p_2} - 1}, \\
W(F_{12}|F_{11}-\lambda_{11} \mathbb{I}) &= \frac{1}{\sqrt{p_2(1-p_1)}} e^{i \varphi_F} \left[(1-\lambda_{11})(\lambda_{11} + p_2 - p_1 p_2 -1)\right]^{1/2}.
\end{align*}
Here $\lambda \in \Omega = [1 - (1-p_1)p_2, 1]$ can be treated as a free parameter. First of the above two sets is a line segment placed on the real axis while the second one forms a circle centered at zero. The nontrivial intersection of these two sets is possible only if $\varphi_F = 0$ and:
\begin{align*}
\lambda_{11} = 1 - \frac{p_2(1-p_1)}{2-p_1-p_2+p_1 p_2}.
\end{align*}
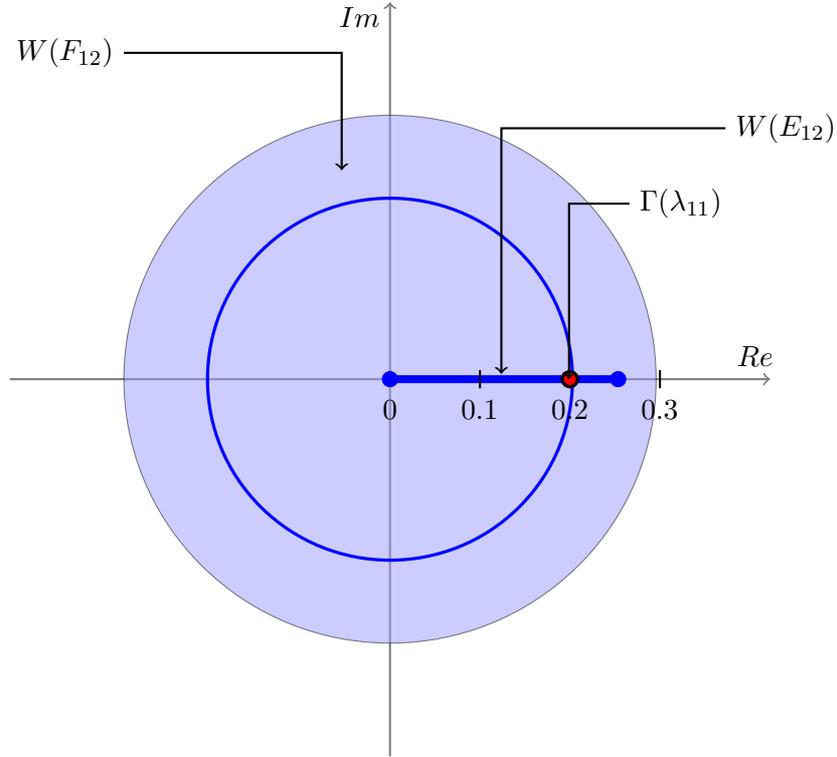
\begin{figure}[!ht]
  \label{fig:5}
  \centering  
  \begin{tikzpicture}[scale=1.]
	\begin{scope}[xshift=10cm]
		\draw[thick,->,black!50] (-5.0,0) -- (5.0,0);
        \draw[thick,->,black!50] (0,-5.0) -- (0,5.0);
	\end{scope}
	\node[xshift=10cm] (X) at (4.8,0.3) {$Re$};
	\node[xshift=10cm] (Y) at (-0.4,4.8) {$Im$};
	
  	\begin{scope}[xshift=10cm, rotate=45]
        \draw[thin, fill=blue!40, opacity = .5] (0,0) circle (3.5);
        \draw[very thick, blue] (0,0) circle (2.4);
    \end{scope}
      
	\draw[thickest,full=0,full=1, blue] (10,0) -- (13,0);    
    
	\node[xshift=10cm, above left] (A) at (-3.5, 4) {$W(F_{12})$};
	\node[xshift=10cm, above left] (B) at (-0.5, 2.5) {};
	
	\node[xshift=10cm, above left] (C) at (+6, 3) {$W(E_{12})$};
	\node[xshift=10cm, above left] (D) at (1.6, -0.2) {};
	
	\draw[xshift=10cm, thick] (3.55, -0.12) -- (3.55, 0.12);
    \node[xshift=10cm] at (3.55,-0.4) {$0.3$};		
	
	\draw[xshift=10cm, thick] (2.366, -0.12) -- (2.366, 0.12);
    \node[xshift=10cm] at (2.366,-0.4) {$0.2$};	
	
	\draw[xshift=10cm, thick] (1.183, -0.12) -- (1.183, 0.12);
    \node[xshift=10cm] at (1.183,-0.4) {$0.1$};
    
	\node[xshift=10cm] at (0.0,-0.4) {$0$};
	\node[xshift=10cm, above left] (E) at (4.5, 2) {$\Gamma (\lambda_{11})$};
	\node[xshift=10cm, above left] (F) at (2.49, -0.27) {};
	\draw[xshift=10cm,very thick, fill = red] (2.366, 0) circle (0.1);
	\draw[->, to path={-| (\tikztotarget)}, thick]
  			(A) edge (B) ;
  	\draw[->, to path={-| (\tikztotarget)}, thick]
  			(C) edge (D) ;
  	\draw[->, to path={-| (\tikztotarget)}, thick]
  			(E) edge (F) ;
	\end{tikzpicture}
  \caption{Standard numerical ranges $W(E_{12})$ (thick line segment on real axis) and $W(F_{12})$ (dark disk) for $p_1 = 0.5$ and $p_2 = 0.7$. Intersection of their respective nuclear numerical ranges  $\Gamma(\lambda_{11}) = W(E_{12}|E_{11}-\lambda_{11} \mathbb{I})  \cap W(F_{12}|F_{11}-\lambda_{11} \mathbb{I}) $ is given by a single point (dot) at approximately $\Gamma(0.696\ldots)\approx(0.2, 0)$.}
\end{figure}
Thus, the error correction code for this specific model is determined by the projector $P_2$ of the form (\ref{eq:21}) with vectors $\Ket{\psi_E} = \Ket{\phi_E}$ and $\Ket{\psi_F} = \Ket{\phi_E}$ defined in Eq. (\ref{eq:17}) with the following horizontal angles:
\begin{align*}
\theta_E &=  \cos^{-1} \left( \frac{p_1 - p_2 + p_1 p_2}{2-p_1-p_2+p_1 p_2} \right),\\
\theta_F &=  \cos^{-1} \left( \frac{-p_1 - p_2 + p_1 p_2}{2-p_1-p_2+p_1 p_2} \right). 
\end{align*}
The polar angle $\varphi_E$ can be chosen arbitrarily: $\varphi_E \in [0, 2 \pi)$, as shown in Fig. 2.

\subsection{General block-diagonal channel of length two}
Let us now consider a more general case of a quantum channel of length two ($k = 2$) acting on system consisting of two qubits:
\begin{align}
\Phi^{G}(\rho) = A_1 \rho A_1^{\dagger} + A_2 \rho A_2^{\dagger}.
\end{align}
Our motivation is to find the most general noise model with a maximal number of free parameters, whose Kraus representation consists of two block-diagonal matrices $\{ A_1, A_2 \}$. Operators $A_i$ in general contain 16 free variables $\{a_1,\, \ldots a_{16}\}$, where $a_i \in \left(0, 1\right)$ and $1 \leq i \leq 16$. The condition that $\Phi^{G}(\rho)$ preserves the trace:
\begin{equation}
\label{eq:ex1}
A_1 A_1^{\dagger} + A_2 A_2^{\dagger} = \mathbb{I}
\end{equation}
imposes additional six constraints (there are eight equations for nonzero block-diagonal elements, from which two are not independent), so that we have 10 free parameters in total. Without loss of generality we can choose any 6 parameters $a_i$ from the set $\{a_1,\, a_2, \ldots a_{16}\}$ and express them in terms of the remaining ones. We label them by $b_j$ for $1 \leq j \leq 6$ so that $\{b_j\}$ are all dependent. Let us label the vector of free parameters by $\vec{a} = (a_1,\, a_2, \ldots a_{10})$. Having this in mind, we can write Kraus operators $A_i$ in the following form:
\begin{align}
\label{eq:ex2}
A_1 = 
\begin{bmatrix}
a_1 & a_2 & 0 & 0 \\
a_3 & a_4 & 0 & 0 \\
0 & 0 & a_5 & a_6 \\
0 & 0 & a_7 & a_8 \\
\end{bmatrix}, \qquad
A_2 = 
\begin{bmatrix}
b_1 & b_2 & 0 & 0 \\
a_9 & b_3 & 0 & 0 \\
0 & 0 & b_4 & b_5 \\
0 & 0 & a_{10} & b_6 \\
\end{bmatrix},&
\end{align}
where the trace-preserving condition (\ref{eq:ex1}) implies:
\begin{align}
\label{eq:ex3}
b_1 &= \sqrt{1-a_1^2-a_3^2-a_9^2}, \quad
b_2 = \frac{\left(a_1 a_2+a_3 a_4\right) b_1 - c_1a_9}{a_1^2+a_3^2-1},  &\\
b_3 &= \frac{\left(a_1 a_2+a_3 a_4\right) a_9+ c_1 b_1}{a_1^2+a_3^2-1}, \quad 
b_4 = \sqrt{1-a_5^2-a_7^2-a_{10}^2}, \nonumber &\\
b_5 &= \frac{\left(a_5 a_6+a_7 a_8\right) b_4-c_2a_{10}}{a_5^2+a_7^2-1}, \quad 
b_6 =\frac{\left(a_5 a_6+a_7 a_8\right) a_{10}+ c_2 b_4}{a_5^2+a_7^2-1}. &\nonumber
\end{align}
To simplify notation we introduce variables $c_i$ which are functions of the independent parameters:
\begin{align}
c_1 &= \sqrt{\left(a_4^2-1\right) a_1^2-2 a_1 a_2 a_3 a_4-a_4^2+\left(a_2^2-1\right) \left(a_3^2-1\right)} & \\ \nonumber
c_2 &= \sqrt{\left(a_8^2-1\right) a_5^2-2 a_5 a_6 a_7 a_8-a_8^2+\left(a_6^2-1\right) \left(a_7^2-1\right)} &
\end{align}
In order to find QECC for the map $\Phi^{G}$ we proceed with the method described in Sect.~\ref{sec:3}. Let us begin by computing matrices $T_{11}$ and $T_{12}$. Defining:
\begin{align*}
e_1 &= a_1^2 + a_3^2, &e_2 = a_1a_2 + a_3a_4, &&e_3 = a_2^2 + a_4^2, & \\
e_4 &= a_1 b_1 + a_3 a_9, &e_5 = a_1 b_2 + a_3 b_3, &&e_6 = a_2 b_1 + a_4a_9, & \\
e_7 &= a_2b_2 + a_4b_3, &f_1 = a_5^2 + a_7^2, &&f_2 = a_5a_6 + a_7a_8, & \\
f_3 &= a_6^2+a_8^2, &f_4 = a_5b_4+a_7a_{10}, &&f_5 = a_5b_5+a_7b_6, & \\
f_6 &= a_6b_4 + a_8a_{10}, &f_7 = a_6b_5 + a_8b_6, &
\end{align*}
one can write matrices $T_{11}$ and $T_{12}$ as:
 \begin{align}
\label{eq:ex4}
T_{11} &= 
\begin{bmatrix}
e_1 & e_2 & 0 & 0 \\
e_2 & e_3 & 0 & 0 \\
0 & 0 & f_1 & f_2 \\
0 & 0 & f_2 & f_3 \\
\end{bmatrix}, \qquad 
T_{12} = 
\begin{bmatrix}
e_4 & e_5 & 0 &  0 \\
e_6 & e_7 & 0 & 0 \\
0 & 0 & f_4 & f_5 \\
0 & 0 & f_5 & f_6 \\
\end{bmatrix}.
\end{align}
Matrices $E_{11}$, $F_{11}$, $E_{12}$ and $F_{12}$ are then:
\begin{align}
E_{11} &= \begin{bmatrix}
e_1 & e_2 \\
e_2 & e_3 
\end{bmatrix}, \quad
F_{11} = 
\begin{bmatrix}
f_1 & f_2 \\
f_2 & f_3
\end{bmatrix}, \\ 
E_{12} &= \begin{bmatrix}
e_4 & e_5 \\
e_6 & e_7
\end{bmatrix}, \quad
F_{12} = \begin{bmatrix}
f_4 & f_5 \\
f_6 & f_7
\end{bmatrix}.
\end{align}
Once again our aim is to find projection operator $P_2$ satisfying Eq. (\ref{eq:3}). We will do it by first computing the intersection of nuclear numerical ranges as explained in Sect.~\ref{sec:3}. To do so we first compute the intersection $\Omega = W(E_{11})\cap W(F_{11})$ which, according to Eq. (\ref{eq:15a}), is completely determined by the eigenvalues of matrices $E_{11}$ and $F_{11}$, denoted by $\nu_{1}$,  $\nu_{2}$ and $\mu_{1}$, $\mu_{2}$, respectively. Without loss of generality we may assume $\nu_{2} > \nu_{1}$ and $\mu_{2} > \mu_{1}$. The set $\Omega$ is then given by:
\begin{align}
\Omega = [\max(\nu_1,\mu_1), \min(\nu_2, \mu_2)].
\end{align}
This set is not empty if condition (\ref{eq:15a}) holds for matrices $E_{11}$ and $F_{11}$. Let us denote by $\lambda_{11}$ a free parameter contained in $\Omega$. In order to find an appropriate QECC one can determine the set $\Gamma(\lambda_{11})$ defined in Eq. (\ref{eq:15c}). We plot the set $\Gamma(\lambda_{11})$ for some convenient choice of parameters $\{ a_i\}$ in Fig. 3.

\begin{figure}[!ht]
  \label{fig:7}
  \centering
 \begin{tikzpicture}
   
	\begin{scope}[xshift=10cm]
		\draw[thick,->,black!50] (-5.0,0) -- (5.0,0);
        \draw[thick,->,black!50] (0,-5.0) -- (0,5.0);
	\end{scope}
	\node[xshift=10cm] (X) at (4.8,0.3) {$Re$};
	\node[xshift=10cm] (Y) at (-0.4,4.8) {$Im$};
	
  	\begin{scope}[xshift=10cm, rotate=45]
        \draw[thin, fill=blue!40, opacity = .5] (.8,1.5) ellipse (2.6 and 1.0);
    \end{scope}
    
    \begin{scope}[xshift=10cm, rotate=-120]     
        \draw[thin, fill=blue!40, opacity = .6] (-1.0,1.6) ellipse (1.0 and 3.0);
    \end{scope}
  
  	\begin{scope}[xshift=10cm, rotate=-120]     
        \draw[very thick,blue] (-1,0.5) ellipse (.6 and 1.6);
    \end{scope}
    
    \begin{scope}[xshift=10cm, rotate=45]
        \draw[very thick, blue] (1.4,1.5) ellipse (1.4 and .7);
    \end{scope}

    \draw[xshift=10cm, thick] (3.55, -0.12) -- (3.55, 0.12);
    \node[xshift=10cm] at (3.55,-0.4) {$0.5$};
    
    \draw[xshift=10cm, thick] (3.55, -0.12) -- (3.55, 0.12);
    \node[xshift=10cm] at (3.55,-0.4) {$0.5$};
    \draw[xshift=10cm, thick] (-0.12, 3.55) -- (0.12, 3.55);
    \node[xshift=10cm] at (-0.4,3.55) {$0.5$};
    
    \node[xshift=10cm] at (-0.2,-0.4) {$0$};
    
    \node[xshift=10cm, above left] (E) at (-2.5, 1.) {$W(E_{12})$};
	\node[xshift=10cm, above left] (Ea) at (-1.5, 0.5) {};
	
	\node[xshift=10cm, above left] (A) at (5.5, 3.5) {$W(E_{12}|E_{11}-\lambda_{11}\mathbb{I})$};
	\node[xshift=10cm, above left] (B) at (.8, 2.91) {};
	
	\node[xshift=10cm] (F) at (2,-2.0) {$W(F_{12})$};
	\node[xshift=10cm] (Fa) at (3,-1.0) {};
	
    \node[xshift=10cm, above left] (C) at (-1.0, -1.2) {$W(F_{12}|F_{11}-\lambda_{11}\mathbb{I})$};
	\node[xshift=10cm, above left] (D) at (2, -0.35) {};
	
	\draw[->, to path={-| (\tikztotarget)}, thick]
  			(F) edge (Fa) ;
  	\draw[->, to path={-| (\tikztotarget)}, thick]
  			(E) edge (Ea) ;
  	\draw[->, to path={-| (\tikztotarget)}, fill = blue, thick]
  			(A) edge (B) ;
  	\draw[->, to path={-| (\tikztotarget)}, fill = blue, thick]
  			(C) edge (D) ;
    
    \begin{scope}[xshift=10cm]
        \draw[very thick, fill = red] (0.37, 1.5) circle (0.1);
        \draw[very thick, fill = red] (-0.37, 1) circle (0.1);
        \draw[very thick, fill = red] (-2.97, 4.32) circle (0.1);
    \end{scope}
    
    \node[xshift=10cm, above left] (O) at (-2.5, 4) {$\Gamma (\lambda_{11})= \{\quad\}$};
    
\end{tikzpicture}
  \caption{Standard numerical ranges $W(E_{12})$ and $W(F_{12})$ (denoted by dark elliptic disks) for the choice of parameter vector $\vec{a} = (a_1,\, a_2, \ldots a_{10}) = (0.9,\ 0.7,\ 0.2,\ 0.9,\ 0.6,\ 0.7,\ 0.9,\ 0.1,\ 0.6,\ 0.5)$. Nuclear numerical ranges in this example are given by two elliptic curves in bold parametrized by $\lambda_{11}$. Their intersection $W(E_{12}|E_{11}-\lambda_{11}\mathbb{I})\cap W(F_{12}|F_{11}-\lambda_{11}\mathbb{I})$ for a given value of $\lambda_{11}$ forms the set $\Gamma(\lambda_{11})$, which in this case consists of two points (dots).}
\end{figure}
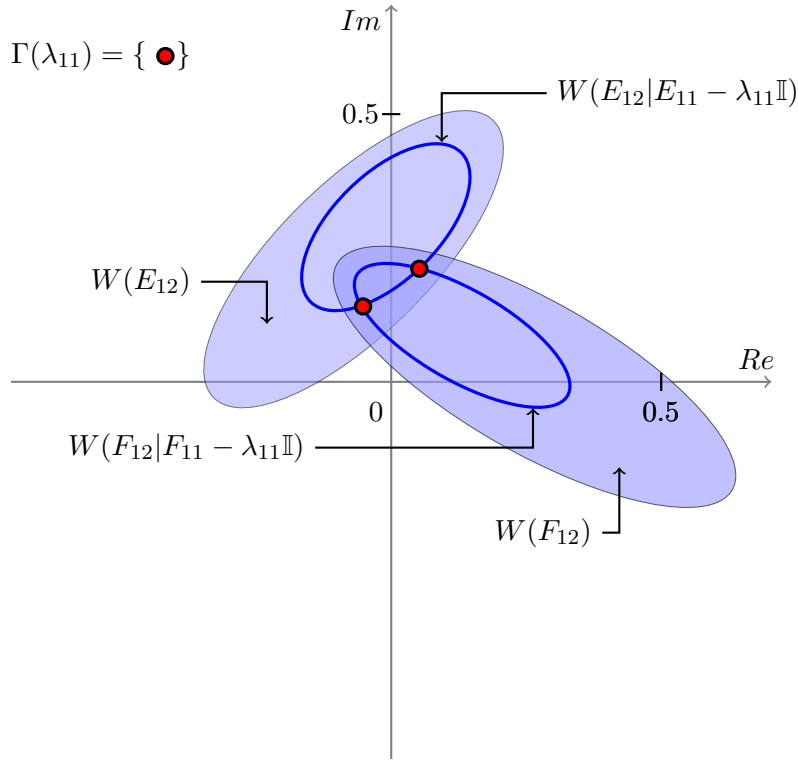

 Following the reasoning from Sect.~\ref{sec:3} and methodology present in the proof of Theorem \ref{t1} we conclude that for a given value of $\lambda_{11}$ one can construct elliptic curves in the complex plane and determine their intersection points $\widetilde{z} = \widetilde{x} + i \widetilde{y} \in \Gamma(\lambda_{11})$ using Theorem $\ref{t1}$. Having obtained the set $\Gamma(\lambda)$ and using Eq. (\ref{eq:18}) and Eq. (\ref{eq:19}) one can then determine the azimuthal and polar angles $(\theta_E,\ \varphi_E)$ for $\Ket{\phi_E}$, and analogous angles $(\theta_F,\ \varphi_F)$ for vector $\Ket{\phi_F}$, respectively. The projection vectors $\Ket{\psi_E}$ and $\Ket{\psi_F}$, which form the projection operator $P_2$, can be computed using Eq. (\ref{eq:20}). The correction code subspace for this particular noise model is then given by Eq. (\ref{eq:21}).

\section{Concluding remarks}
\label{sec:5}
In this work we have introduced the notion of nuclear numerical range $W(A|Z)$ of an operator $A$ with respect to an auxiliary operator $Z$ and demonstrated that it allows one to find quantum error-correction codes protecting against noise. In particular, this technique works for models of quantum errors with non-unitary noise operators. Using a simple geometric construction involving an intersection of two ellipses in the complex plane we found such a quantum error-correction code for a simplified model of two-qubit amplitude damping channel and a general noise model with two Kraus operators of size $n = 4$ with block-diagonal structure. Note that the method used here for the two-qubit system is straightforward to generalize for larger dimensions. We expect that further development of this technique will allow for effective construction of quantum error-correction codes protecting information against more general non-unitary noise models.
\section*{Appendix A}
\label{app}
In this Appendix a proof of Theorem \ref{t1} is presented.
\begin{proof}
Consider:
\begin{align*}
Z &= 
\begin{bmatrix}
2 a & b \\
b & 2 c
\end{bmatrix} \in \mathcal{M}_2^{\mathbb{R}}, \quad
A =
\begin{bmatrix}
d & f \\
g & h
\end{bmatrix} \in \mathcal{M}_2^{\mathbb{C}}.
\end{align*}
Let us start by diagonalizing the matrix $Z$. To do so, we first subtract half of the trace from the diagonal to obtain a traceless matrix $Z' = Z - \left(\frac{1}{2} \Tr Z\right) \mathbb{I}$ with eigenvalues $\pm \epsilon \equiv \pm \left[b^2 + (a-c)^2\right]^{1/2}$. Since $Z'$ is real and symmetric it can be diagonalized by an orthogonal matrix \cite{horn}:
\begin{align}
\label{leq:1}
U = \begin{bmatrix}
\cos \alpha & \sin \alpha \\
\sin \alpha & - \cos \alpha
\end{bmatrix},
\end{align}
where:
\begin{align}
\label{leq:1aa}
\tan 2\alpha = b/(a-c),
\end{align} and $U Z' U^{\dagger} = \diag(\epsilon, -\epsilon)$. Let us introduce a unit vector: 
\begin{align}
\label{leq:1a}
\Ket{\phi}=
\begin{pmatrix}
\cos\frac{\theta}{2} \\ \text{e}^{i \varphi}\sin\frac{\theta}{2}
\end{pmatrix},
\end{align}
with $0 \leq \varphi < 2\pi$ and $0 \leq \theta < \pi$ and such that $\Ket{\psi} = U\Ket{\phi}$, where $\Ket{\psi}$ is some unit vector in $\mathbb{C}^2$. Having this in mind, the first condition from Eq. (\ref{teq:1}) becomes:
\begin{align*}
\Braket{\psi|Z|\psi} = \lambda = \Braket{\phi|U^{\dagger} Z' U|\phi} + \frac{1}{2} \Tr Z = \epsilon \cos \theta + \frac{1}{2} \Tr Z,
\end{align*} 
where $\pm \epsilon$ are the eigenvalues of $Z$ defined in text above Eq. (\ref{leq:1}).  Since $0 \leq \theta < \pi$, we also have $\lambda \in \Gamma_{\epsilon} = [\frac{1}{2} \Tr Z-\epsilon, \frac{1}{2} \Tr Z + \epsilon]$. Thus, assuming that $Z$ is non-degenerate, we can express the parametrization angle $\theta$ in terms of the compression value $\lambda$ as: 
\begin{align}
\label{leq:2}
\cos \theta = \frac{1}{\epsilon}\left(\lambda - \frac{1}{2} \Tr Z\right).
\end{align}
Consider now the second Eq. from (\ref{teq:1}). In our current parametrization it reads:
\begin{align*}
\Braket{\psi|A|\psi} = \frac{1}{2} \Tr A + \sin 2 \alpha \left[ \frac{f+g}{2} \cos \theta + \frac{d-h}{2} \sin \theta \cos \varphi \right] + \\ \cos 2 \alpha \left[ \frac{d-h}{2} \cos \theta - \frac{f+g}{2} \sin \theta \cos \varphi \right] - i \frac{f-g}{2} \sin \varphi.
\end{align*}
By substituting $\tan 2\alpha = b/(a-c)$ into above expression and using Eq. (\ref{leq:2}) we obtain:
\begin{multline}
\label{leq:3}
\Braket{\psi|A|\psi} = \frac{d+h}{2} + \frac{\sgn (a-c)}{2 \epsilon^2} \bigg \{(\lambda - a - c) \left[b(f+g)+(d-h)(a-c)\right] \\ 
 +\left[b(d-h)-(f+g)(a-c)\right] \left[\epsilon^2 - (a+c-\lambda)^2\right]^{1/2} \cos  \varphi \\- i (f-g) \sgn (a-c) \epsilon \sin \varphi \bigg \}.
\end{multline}
Let us simplify the notation by denoting:
\begin{equation}
\label{leq:4}
\begin{aligned}[c]
z_0 & = \frac{d+h}{2} - \frac{\sgn (a-c)}{2\epsilon^2} \Big \{ (a+c)[b(f+g)+(a-c)(d-h)]\Big \},\\
w & = \frac{\sgn (a-c)}{2 \epsilon^2}\left[b(f+g) + (a-c)(d-h)\right], \\ 
p(\lambda) &= \frac{\sgn (a-c)}{2 \epsilon^2 }\left[\epsilon^2 - (a-c-\lambda)^2\right]^{1/2}, \\
q & = b(d-h)-(a-c)(f+g), \quad r = -i \epsilon (f-g) \sgn (a-c).
\end{aligned}
\end{equation}
Using the above we can rewrite Eq. (\ref{leq:3}) in the following form:
 \begin{align}
 \label{leq:5}
\Braket{\psi|A|\psi} =  z(\varphi, \lambda) = z_0 + w \lambda + p(\lambda) [q \cos \varphi + r \sin \varphi].
 \end{align} 
Eq. (\ref{leq:5}) defines an entire family of ellipses in the complex plane which belong to $W(A)$. In particular, if we let $\lambda$ to run over its available range, that is $\Gamma_{\epsilon} = [\frac{1}{2} \Tr Z-\epsilon, \frac{1}{2} \Tr Z + \epsilon]$, we recover the standard numerical range $W(A)$. To show this, let us choose a coordinate frame in which position is parametrized by $\varphi \in [0, 2\pi)$ and $\lambda \in \Gamma_{\epsilon}$, and is given by a complex number $z(\varphi, \lambda) = z_0 + w\lambda + p(\lambda)\left[x(\varphi) + iy(\varphi)\right]$, where $x(\varphi),y(\varphi) \in\mathbb{R}$. Let us denote $q_1 = \Re(q)$, $q_2 = \Im(q)$, $r_1 = \Re(r)$, $r_2 = \Im(r)$, $x_0(\lambda) = \Re(z_0 + w\lambda)$, $y_0(\lambda) = \Im( z_0 + w\lambda)$ and choose $x(\varphi)$ and $y(\varphi)$ in the following way:
\begin{align}
\label{leq:6a}
x(\varphi) &= q_1 \cos \varphi + r_1 \sin \varphi, \\ 
\label{leq:6b}
y(\varphi) &= q_2 \cos \varphi + r_2 \sin \varphi.
\end{align}
Recall that the equation of an ellipse centered at zero is given by:
\begin{align}
\label{leq:7}
\alpha x^2 + \beta y^2 + \gamma xy = 1, \\ \text{given that} \qquad \gamma^2 - 4 \alpha \beta < 0.
\end{align}
Let us now plug-in the coordinates $x(\varphi)$ and $y(\varphi)$ into above equation. Making use of simple trigonometric identities the above equation of the ellipse can be rewritten as:
\begin{align}
\label{leq:8}
&\left[\alpha (q_1^2 + r_1^2) + \beta (q_2^2 + r_2^2) + \gamma (q_1 q_2 + r_1 r_2) \right] \nonumber \\ 
+ &\left[\alpha(q_1^2 - r_1^2) + \beta (q_2^2 - r_2^2) + \gamma (q_1 q_2 - r_1 r_2)  \right] \cos 2 \varphi \nonumber \\ 
+ &\left[2 q_1 r_1 \alpha + 2 q_2 r_2 \beta + \gamma (q_2 r_1 + q_1 r_2) \right]\sin 2 \varphi = 2p^2.
\end{align}
This can be satisfied for all values of $\varphi$ only if the following set of equations is satisfied:
\begin{align}
\begin{cases}
\alpha (q_1^2 + r_1^2) + \beta (q_2^2 + r_2^2) + \gamma (q_1 q_2 + r_1 r_2) = 2p^2 \\ \nonumber
\alpha(q_1^2 - r_1^2) + \beta (q_2^2 - r_2^2) + \gamma (q_1 q_2 - r_1 r_2) = 0 \\ \nonumber
2 q_1 r_1 \alpha + 2 q_2 r_2 \beta + \gamma (q_2 r_1 + q_1 r_2)  = 0.
\end{cases}
\end{align}
By solving the above set of equations we can easly determine coefficients $\alpha, \beta$ and $\gamma$,
\begin{align}
\label{eq:ecoefs}
\alpha =& \frac{q_2^2 + r_2^2}{(r_1q_2 - q_1 r_2)^2}, \nonumber \\
\beta =& \frac{q_1^2 + r_1^2}{(r_1q_2 - q_1 r_2)^2}, \\
\gamma =& -2 \frac{q_1 q_2 + r_1 r_2}{(r_1q_2 - q_1 r_2)^2}. \nonumber
\end{align}
A simple calculation shows that discriminant of Eq. (\ref{leq:7}) is negative:
\begin{align*}
\gamma^2 - 4 \alpha \beta = -\frac{4}{(q_2r_1-q_1r_2)^2} < 0,
\end{align*}
since all variables in above expression are real. If we now rescale $x(\varphi)$ and $y(\varphi)$ by a factor of $p(\lambda)$ and shift it accordingly by $x_0(\lambda)$ and $y_0(\lambda)$ so that $\widetilde{x}(\varphi, \lambda) = x_0(\lambda) + p(\lambda) x(\varphi)$ and $\widetilde{y}(\varphi, \lambda) = y_0(\lambda) + p(\lambda) y(\varphi)$, we get $\widetilde{z}(\varphi, \lambda) = \widetilde{x}(\varphi, \lambda) + i \widetilde{y}(\varphi, \lambda) = z(\varphi, \lambda)$. Since an affine transformation sends an ellipse to an ellipse, also $z(\varphi, \lambda)$ describes an ellipse. This completes the proof that $z(\lambda, \varphi)$ defines an ellipse. In order to determine states $\Ket{\psi} = U \Ket{\phi}$, where $\Ket{\phi}$ is given by Eq. (\ref{leq:1a}), one has to find angles $\theta, \varphi, \alpha$ in terms of the point $\widetilde{z} = \widetilde{x} + i \widetilde{y}$ in the complex plane and a real parameter $\lambda$. From Eq. (\ref{leq:2}) one determines the azimuthal angle $\theta$ in terms of $\lambda$. Using the definition of $\widetilde{z} = \widetilde{z}(\varphi, \lambda)$ one then finds:
\begin{align*}
x(\varphi) = \frac{\widetilde{x} - x_0(\lambda)}{p(\lambda)} \quad \text{and} \quad y(\varphi) = \frac{\widetilde{y} - y_0(\lambda)}{p(\lambda)}.
\end{align*}
Using above expressions and Eqs. (\ref{leq:6a}) and (\ref{leq:6b}) one can determine the polar angle $\varphi$:
\begin{align*}
\cos \varphi = \frac{x r_2 - y r_1}{q_1 r_2 - r_1 q_2} = \frac{1}{p(\lambda)}\frac{r_2\left[\widetilde{x} - x_0(\lambda)\right] - r_1\left[\widetilde{y} - y_0(\lambda))\right]}{q_1 r_2 - r_1 q_2}.
\end{align*} 
The angle $\alpha$ can be computed by recalling that $\tan 2\alpha = b / (a-c)$.
The family of states $\Ket{\psi}$ is then given by
\begin{align*}
\Ket{\psi} = 
\begingroup
\renewcommand*{\arraystretch}{1.3}
\begin{pmatrix}
\cos \alpha \cos \frac{\theta}{2} + \text{e}^{i \varphi} \sin \alpha \sin \frac{\theta}{2} \\ 
\sin \alpha \cos \frac{\theta}{2} + \text{e}^{i \varphi} \cos \alpha \sin \frac{\theta}{2}
\end{pmatrix},
\endgroup
\end{align*}
which completes the proof of the second part of the theorem.
\end{proof}
\section*{Acknowledgements}
We are grateful to the anonymous Referee for several constructive comments which allowed us to improve our work. This work has been supported by the Polish National Science Center under the project number DEC-2015/18/A/ST2/00274 and by the John Templeton Foundation under the project No. 56033


\end{document}